\documentclass[twoside,fleqn]{article}
\usepackage{espcrc2}
\usepackage{theorem}
\usepackage{epsfig}
\usepackage{graphicx}

\theoremstyle{break}    
\theoremstyle{plain}    
\theoremstyle{plain}    
\theoremstyle{plain}    
{\theorembodyfont{\rmfamily}     }
{\theorembodyfont{\rmfamily}     }

\def\gsim{{\mathrel{\raise2pt\hbox to 8pt{\raise -5pt\hbox{$\sim$}\hss{$>$}}}}}
\def\rsim{{\mathrel{\raise2pt\hbox to 8pt{\raise -5pt\hbox{$\sim$}\hss{$>$}}}}}
\def\lsim{{\mathrel{\raise2pt\hbox to 8pt{\raise -5pt\hbox{$\sim$}\hss{$<$}}}}}

\begin{document}

\title{
$Re(A_0)$, $Re(A_2)$ and RG evolution for $N_f=3$\thanks{
Presented by W.~Lee.  Research supported in part by BK21, by
Interdisciplinary Research Grant of Seoul National University, and by
KOSEF contract R01-2003-000-10229-0.}  }
\author{Keunsu Choi\address[SNU]{School of Physics, 
		Seoul National University,
		Seoul, 151-747, South Korea}
        and
        Weonjong Lee\addressmark[SNU]
}
\begin{abstract}
We present results of $Re (A_0)$ and $Re (A_2)$ calculated using HYP
staggered fermions on the lattice of $16^3 \times 64$ at $\beta=6.0$.
These results are obtained using leading order chiral perturbation in
quenched QCD. Buras's original RG evolution matrix develops a
removable singularity for $N_f=3$. This subtlety is resolved by
finding a finite solution to RG equation and the results are
presented.
\end{abstract}

\maketitle

\begin{figure}[t!]
\epsfig{file=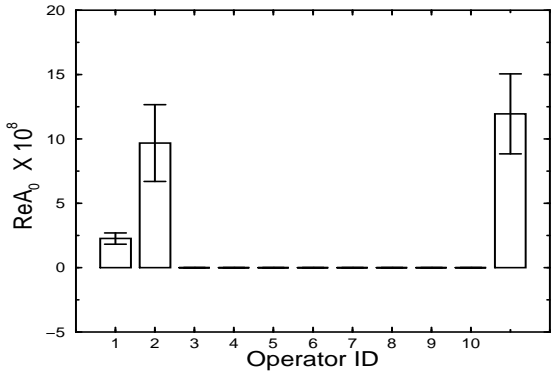, height=12pc, width=17pc}
\vspace*{-5mm}
\caption{$Re(A_0)$, linear fitting for chiral extrapolation,
combined at $\mu=m_c$.}
\label{fig:ReA0:lin:mc}
\vspace*{-4mm}
\end{figure}

In the standard model, we can express $Re(A_I)$ in terms of the matrix
elements of the effective weak Hamiltonian between hadronic states as
follows:
\begin{eqnarray}
    Re (A_I) &=& \frac{G_F}{\sqrt{2}} |V_{ud} V_{us}|
    \Big[ \sum_{i=1,2} z_i (\mu) \langle Q_i (\mu) \rangle_I 
\nonumber \\ & & 
      + Re(\tau) \sum_{i=3}^{10} y_i (\mu) \langle Q_i(\mu) \rangle_I
      \Big]
\label{eq:Re(A_I)}
\end{eqnarray}
where $V_{ij}$ is an element of CKM matrix.
Here, $z_i(\mu)$ and $y_i(\mu)$ are the Wilson coefficients which are
obtained analytically.
The hadronic matrix elements $\langle Q_i(\mu) \rangle_I$ are
defined as
\begin{equation}
\langle Q_i(\mu) \rangle_I = \langle \pi\pi_I | Q_i(\mu) | K \rangle
\end{equation}
Since these matrix elements are in a highly non-perturbative regime of
QCD, it is necessary to calculate them using non-perturbative tools
such as a numerical method based on lattice gauge theory.
In fact, we calculate $ \langle \pi | Q_i | K \rangle $ and $ \langle
0 | Q_i | K \rangle $ on the lattice using HYP staggered fermions and
we construct $ \langle \pi\pi_I | Q_i | K \rangle $ out of them using
chiral perturbation in its leading order \cite{ref:soni:1}.
The $Re(\tau)$ in Eq.~(\ref{eq:Re(A_I)}) is defined as
\begin{eqnarray*}
Re(\tau) &=& - Re(\lambda_t/\lambda_u) = 0.002 \\
\lambda_f &=& V_{fs}^{*} V_{fd}
\end{eqnarray*}
In Eq.~(\ref{eq:Re(A_I)}), $ Q_i $ ($i=3,4,\cdots,10$) represents the
QCD and electroweak penguin operators.
Since $z_i, y_i \approx 1$, the contribution from the penguin
operators to $Re(A_I)$ is suppressed by $Re(\tau)$.
Therefore, in practice, only the current-current operators $\langle
Q_i \rangle$ ($i=1,2$) can contribute dominantly to $Re(A_I)$ whereas
the contribution from penguin operators is negligible.
In the case of the direct CP violation, only penguin operators can
contribute to $\epsilon'/\epsilon$ \cite{ref:wlee:1} and
current-current operators do not play any role in it.
In this regards, we can say that $\epsilon'/\epsilon$ and $Re(A_I)$
are completely different physics.
In Fig.~\ref{fig:ReA0:lin:mc}, we show individual channels
contributing to $Re(A_0)$ with the total sum in the last column.
The dominant contribution comes from $\langle Q_i \rangle$ $(i=1,2)$ as
expected.
Each channel is obtained by fitting the data to a function: $f_1(m_K)
= c_0 + c_1 m_K^2$ and extrapolating to the chiral limit.
We use the matching formula given in \cite{ref:wlee:2,ref:wlee:3} to
convert lattice results into continuum values in the NDR scheme at the
scale of $\mu = 1/a$.
By assuming that these values are close enough to those for $N_f=3$,
we run them from $\mu=1/a$ down to $\mu=m_c$, using the RG evolution
equation for $N_f=3$.
There are two alternative methods to transcribe the $Q_6$ and $Q_5$
operators to the lattice: the standard (STD) method and the
Golterman \& Pallante (GP) method \cite{ref:golterman:1}.
In the case of $\epsilon'/\epsilon$, the dominant contribution to the
$\Delta I = 1/2$ amplitude comes from $\langle Q_6 \rangle$ and so
there is a big difference in $\epsilon'/\epsilon$ between the STD and
GP methods \cite{ref:wlee:1}.
In the case of $Re(A_0)$, however, the contribution from $\langle Q_6
\rangle$ is extremely suppressed by $Re(\tau)$ and so both methods do
not make much difference to $Re(A_0)$.
%
%

\begin{figure}[t!]
\epsfig{file=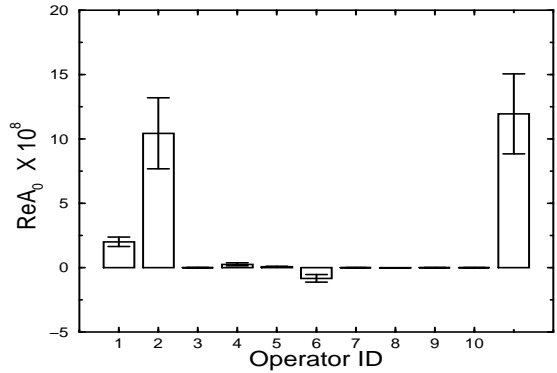, height=12pc, width=17pc}
\vspace*{-5mm}
\caption{$Re(A_0)$, linear fitting for chiral extrapolation,
combined at $\mu=1/a$.}
\label{fig:ReA0:lin:1/a}
\vspace*{-4mm}
\end{figure}

In Fig.~\ref{fig:ReA0:lin:1/a} we show the same kind of plot as in
Fig.~\ref{fig:ReA0:lin:mc} except for the scale at which the Wilson
coefficient and the matrix elements are combined.
We set the scale to $\mu=1/a$ in Fig.~\ref{fig:ReA0:lin:1/a} and to
$\mu=m_c$ (charm quark mass) in Fig.~\ref{fig:ReA0:lin:mc}.
The contribution from each individual channel can fluctuate depending
on this matching scale but the total sum should be invariant.
%

%
\begin{figure}[t!]
\epsfig{file=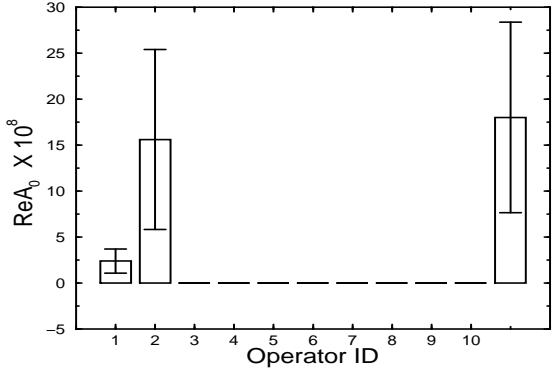, height=12pc, width=17pc}
\vspace*{-5mm}
\caption{$Re(A_0)$, quadratic fitting for chiral extrapolation,
combined at $\mu=m_c$.}
\label{fig:ReA0:quad:mc}
\vspace*{-3mm}
\end{figure}

\begin{figure}[t!]
\epsfig{file=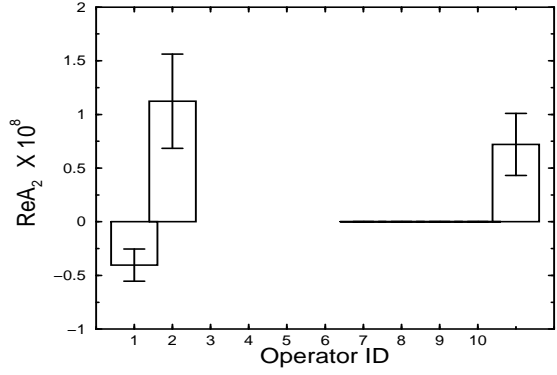, height=12pc, width=17pc}
\vspace*{-5mm}
\caption{$Re(A_2)$, linear fitting for chiral extrapolation,
combined at $\mu=m_c$.}
\label{fig:ReA2:lin:mc}
\vspace*{-3mm}
\end{figure}

In Fig.~\ref{fig:ReA0:quad:mc}, we show the same kind of plot as
in Fig.~\ref{fig:ReA0:lin:mc} except for the fitting function.
Each channel in Fig.~\ref{fig:ReA0:quad:mc} is obtained by fitting the
data to a function: $ f_2(m_K) = c_0 + c_1 m_K^2 + c_2 m_K^4$ and
extrapolating it to the chiral limit.
Comparing Fig.~\ref{fig:ReA0:lin:mc} and \ref{fig:ReA0:quad:mc}, we
observe that the dependence of the chiral extrapolation on the fitting
function is large.
These results are consistent with the experimental value of $Re(A_0) =
33.3 \times 10^{-8}$ GeV within the systematic and statistical
uncertainty.
In Fig.~\ref{fig:ReA2:lin:mc}, we present individual channels
contributing to $Re(A_2)$ and the total sum in the last column.
Each channel is obtained by fitting the data to a linear function
$f_1(m_K)$ and extrapolating it to the chiral limit.
The result of Fig.~\ref{fig:ReA2:lin:mc} is slightly lower than
the experimental value of $Re(A_2) = 1.5 \times 10^{-8}$ GeV.
%

%
%
Now, let us address a subtle issue on the RG evolution for $N_f=3$.
The $\Delta S = 1$ effective Hamiltonian can be written as follows:
\begin{equation}
{\cal H} = \frac{G_F}{\sqrt{2}} \sum_{i=1}^{10} C_i(\mu) Q_i(\mu)
\end{equation}
The RG (renormalization group) equation for the Wilson coefficient $C_i$
is
\begin{equation}
\left[ \mu\frac{\partial}{\partial\mu}
+ \beta(g) \frac{\partial}{\partial g} \right] \vec{C}
= \gamma^T(g, \alpha) \vec{C}
\end{equation}
where $\beta(g)$ is the QCD beta function:
\begin{equation}
\beta(g) = -\beta_0 \frac{g^3}{16\pi^2}
-\beta_1 \frac{g^5}{(16\pi^2)^2} 
-\beta_{1e} \frac{e^2 g^3}{(16\pi^2)^2}
\end{equation}
and $\gamma(g,\alpha)$ is the anomalous dimension matrix:
\begin{eqnarray}
\gamma(g,\alpha) &=& \gamma_s(g^2) + \frac{\alpha}{4\pi} \Gamma(g^2) + \cdots
\\
\gamma_s(g^2) &=& \gamma_s^{(0)} \frac{\alpha_s}{4\pi}  
+ \gamma_s^{(1)} \frac{\alpha_s^2}{ (4\pi)^2 } + \cdots
\end{eqnarray}
A solution to the RG equation can be expressed in terms of the evolution
matrix.
\begin{equation}
\vec{C}(\mu) = U(\mu,\mu',\alpha) \vec{C}(\mu')
\end{equation}
Here the RG evolution matrix is, in general,
\begin{equation}
U(m_1,m_2,\alpha) = T_g \exp\left(
\int^{g(m_1)}_{g(m_2)} dg' \frac{\gamma^T(g',\alpha)}{\beta(g')}
\right)
\end{equation}
In the perturbative expansion, we can express the RG evolution
matrix as follows:
\begin{equation}
U(m_1,m_2,\alpha) = U(m_1,m_2) + \frac{\alpha}{4\pi} R(m_1,m_2)
\end{equation}
where $U(m_1,m_2)$ represents the pure QCD evolution and $R(m_1,m_2)$
describes the additional evolution in the presence of the
electromagnetic interaction.
The pure QCD evolution matrix which Buras, {\em et al.} provided
originally \cite{ref:buras:1,ref:buras:2}, is 
\begin{eqnarray}
U(m_1,m_2) &=& \left( 1 + \frac{\alpha_s(m_1)}{4\pi} J \right)
U^{(0)}(m_1,m_2) 
\nonumber \\ & & \cdot
\left( 1 - \frac{\alpha_s(m_2)}{4\pi} J \right)
\end{eqnarray}
where $U^{(0)}(m_1,m_2)$ denotes the evolution in the
leading logarithmic approximation and $J$ corresponds to
the next-to-leading correction.
%
%
%
The $J$ matrix can be expressed as follows:
\begin{eqnarray}
J &=& V S V^{-1}\,, \quad G = V^{-1} \gamma_s^{(1)T} V \\
S_{ij} &=& \delta_{ij} \gamma_i^{(0)} \frac{\beta_1}{2 \beta_0^2}
- \frac{ G_{ij} }{ 2 \beta_0 + \gamma^{(0)}_i - \gamma^{(0)}_j }
\end{eqnarray}
where $V$ is a matrix which diagonalize the $\gamma^{(0)T}_s$ matrix
and $\gamma^{(0)}_i$ represents corresponding diagonal elements.
Here, note that the $J$ or $S$ matrix is singular if 
$ 2 \beta_0 + \gamma^{(0)}_i - \gamma^{(0)}_j = 0 $.
In fact, for $N_f = 3$, $2 \beta_0 + \gamma^{(0)}_i - \gamma^{(0)}_j =
0$ when $i=8$, $j=7$.
However, since $U(m_1,m_2)$ is not singular, this singularity must be
removable at the next-to-leading order.
After we make a correct combination at the next leading order and
remove the singularity, we can make the $U(m_1,m_2)$ finite. 
\begin{equation}
U(m_1,m_2) = U_0(m_1,m_2) + \frac{1}{4\pi} V A(m_1,m_2) V^{-1}
\end{equation}
where, if $ a_j \ne a_i+1 $ (note that $a_i \equiv \gamma^{(0)}_i/ (2
\beta_0)$),
\begin{eqnarray}
A(m_1,m_2) &=& S_{ij} \biggl[
\alpha_s(m_1) \bigg( \frac{\alpha_s(m_2)}{\alpha_s(m_1)} \bigg)^{a_j}
\nonumber \\ & &
- \alpha_s(m_2) \bigg( \frac{\alpha_s(m_2)}{\alpha_s(m_1)} \bigg)^{a_i}
\biggl]
\end{eqnarray}
and if $ a_j = a_i+1 $,
\begin{eqnarray}
A(m_1,m_2) &=& \frac{G_{ij}}{ 2 \beta_0 } \alpha_s(m_2) \left(
\frac{\alpha_s(m_2)}{\alpha_s(m_1)} \right)^{a_i} 
\nonumber \\ & &
\cdot \ln \left( \frac{\alpha_s(m_2)}{\alpha_s(m_1)} \right)
\end{eqnarray}
Here, note that the $A$ matrix is finite while $J$ matrix is singular.

In the case of the $R(m_1,m_2)$, the same kind of removable singularity
makes the RG evolution much more complicated, which we will present
in Ref.~\cite{ref:wlee:4}.

\end{document}